\documentstyle[aps]{revtex}
\draft
\begin{document}
\title{Pulsar motions from neutrino oscillations induced by a
violation of the equivalence principle}
\author{M. Barkovich$^{\dagger }$, H. Casini$^{*}$,
J.C. D'Olivo$^{\dagger }$, R.Montemayor$^{\ddagger }$}
\address{$^{\dagger }$Instituto de Ciencias Nucleares,
UNAM, Ap. Postal 70-543, 04510\\
M\'{e}xico DF, Mexico\\
$^{*}$ Theoretical Physics, University of Oxford, 1 Keble Road, Oxford\\
OX13NP, United Kingdom\\
$^{\ddagger}$ Instituto Balseiro and CAB, Universidad Nacional de
Cuyo and CNEA,\\ 8400 Bariloche, Argentina}

\maketitle

\begin{abstract}
We analize a possible explanation of the pulsar motions in terms of resonant
neutrino transitions induced by a violation of the equivalence principle
(VEP). Our approach, based on a parametrized post-Newtonian (PPN) expansion,
shows that VEP effects give rise to highly directional contributions to the
neutrino oscillation length. These terms induce anisotropies in the linear
and angular momentum of the emitted neutrinos, which can account for both
the observed translational and rotational pulsar motions. The violation
needed to produce the actual motions is completely compatible with the
existing bounds.
\end{abstract}

\pacs{PACS numbers: 97.60.G, 04.80.Cc, 14.60.Pq, 95.30.Sf}

It is very difficult to obtain precise evidence on the characteristics of the
gravitational interaction beyond the range where the Newtonian approximation
holds. Only systems with very large densities of mass in rapid motion can
provide suitable laboratories for such a phenomenology. One well known
example is the orbital behavior of binary pulsars, which gives support to
the production of gravitational waves. Type II supernovas are another
interesting scenario. In this case the intense neutrino flux produced during
the gravitational collapse can be sensible to subtle characteristics of the
gravitational interaction. In this letter we analyze some effects on this
flux that could test a possible violation of the equivalence principle.

Perhaps one of the most intringuing characteristic of the pulsar dynamics
related with the supernova stage is their anomalous proper motions. There is
strong observational evidence that translational velocities of pulsars
include a significant component from kicks given when they are formed
\cite{velocidad}. Several mechanisms have been proposed to explain such
kicks, but none of them is completely
satisfactory\cite{mecanismos,kusegre,horvat}. Recently it has also been
pointed out that the observed rotation periods are several orders of
magnitude shorter than the predictions for the cores of the protoneutron
stars\cite {spruit}. Thus the spin of the pulsars are probably produced by
the same mechanism that gives them their translational velocities during the
formation stage. Moreover, there is significant observational evidence that
seems to indicate a polarization of the motion of young pulsars along a
direction near the plane of the galaxy\cite{polarization}. This correlation
could mean that kicks involve a characteristic length at least of the order
of the galaxy radius, which is very difficult to explain on the basis of the
proposed mechanisms.

An appealing possibility that could account for the translational kick is a
1\% anisotropy in the momentum carried by the neutrinos emitted during
pulsar formation. However small, such anisotropy is not easy to obtain.
Kusenko and Segre (KS)\cite{kusegre} have proposed a mechanism based on the
deformation of the resonance surface when neutrinos undergo matter
oscillations in the presence of a magnetic field\cite{magnetic}.
Unfortunately the necessary magnetic field is relatively high, $B\gtrsim
10^{15}$ G\cite{kusegre,qian,raffelt}. Furthermore, the condition that the
resonance surface has to lie between the neutrinoespheres implies $m_{\nu
}\sim 100$ eV. The existence of such heavy neutrinos is cosmologically ruled
out unless they are unstable.

A less orthodox mechanism for neutrino oscillation was proposed several years
ago. It requires a flavor dependent coupling of neutrinos to
gravity\cite{gasperini}, and no neutrino mass. Consequences of such a
violation of the equivalence principle (VEP) in the neutrino sector have
been analyzed in a number of papers\cite{vep}. In particular, in
Ref.\cite{horvat} it was applied to the problem of the translational motion
of pulsars . In this case the desired kick can be achieved with massless
neutrinos, but the intensity of the magnetic field is similar to the one
required by the KS mechanism.

In this work we propose a purely gravitational explanation for both the
translational and rotational motion of pulsars, where the neutrino
oscillation and the momentum anisotropy are induced by VEP effects and that
do not rely on the magnetic field of the protostar.  We work within the
framework of a generalized parametrized post-Newtonian (PPN) formalism
\cite{will}, previously applied to the solar neutrino problem\cite{nuestro},
that naturally includes the effect of a preferred reference system. Our
approach generalizes the usual VEP scheme\cite{vep}, by including the effect
of potentials of the next PPN order to the Newtonian potential $U$, and a
tensorial potential of the same order than $U$. In principle all these terms
should be present if the equivalence principle is violated. In this context
the neutrino oscillations are a manifestation of a VEP effect, and the
momentum anisotropy is signature of the preferred reference system. The
accuracy of the equivalence principle may be characterized by limits in the
differences of the PPN parameters for different neutrinos. As we show,
violations of the equivalence principle consistent with the present bounds
generate the necessary kicks to produce the observed pulsar motions.

The linearized Dirac equation for massless neutrinos in a static
gravitational field leads to the dispersion relation\cite{cm}:
\begin{equation}
E=p\left[ 1+h_{oi}\,\hat{p}_{i}\,-\frac{1}{2}h_{ij}\hat{p}_{i}\hat{p}_{j}-
\frac{1}{2}h_{oo}U\right] \,,  \label{HAM}
\end{equation}
where the $h^{\mu \nu }$ fields are defined by $g^{\mu \nu }=\eta ^{\mu \nu
}+h^{\mu \nu }$, referred to the Minkowskian metric. In deriving this
relation we have neglected the spatial derivatives of the gravitational
potentials, which is justified for neutrinos in astrophysical systems. Up to
third order in the velocity of the source $w$ we can write ($G=\hbar =c=1$):
\begin{eqnarray}
h_{oo} &=&2\gamma ^{\prime }U+{\cal O}(w^{4})\,, \\
h_{oi} &=&-\frac{7}{2}\Delta _{1}V_{i}-\frac{1}{2}\Delta _{2}W_{i}+(\alpha
_{2}-\frac{1}{2}\alpha _{1})v_{i}U-\alpha _{2}v_{j}U_{ji}+{\cal O}(w^{4})\,,
\label{PPN} \\
h_{ij} &=&2\gamma U\delta _{ij}+\Gamma U_{ij}\,+{\cal O}(w^{4})\,.
\end{eqnarray}
The adimensional parameters of the PPN expansion are $\gamma $, $\gamma
^{\prime }$, $\Delta _{1}$, $\Delta _{2}$, $\Gamma $, ${\bf v}$ , $\alpha
_{1}$, and $\alpha _{2}$. The parameters $\alpha _{1}$ and $\alpha _{2}$
vanish in Lorentz covariant theories, but if there exists a preferred
reference frame, characterized by a velocity ${\bf v}$, they should be non
null.

The general expressions for the potentials $U$, $V_{i}$, $W_{i}$ and $U_{ji}$
can be found in Ref.\cite{nuestro}. In the present case, the source of the
gravitational field is the protoneutron star. Considering a spherical
configuration and a rigid rotation, the PPN potentials become
\begin{eqnarray}
U &=&4\pi \int_{0}^{R}dr^{\prime }r^{\prime 2}\left[ \frac{1}{r}\theta
\left( r-r^{\prime }\right) +\frac{1}{r^{\prime }}\theta \left( r^{\prime
}-r\right) \right] \rho \left( r^{\prime }\right) \,, \\
U_{ij} &=&\hat{r}_{i}\hat{r}_{j}I(r)+\delta _{ij}\;J(r)\,, \\
V_{i}\; &=&W_{i}\,=\;w_{i}J(r)\,.
\end{eqnarray}
where $\rho (r)$ is the mass distribution of the star and $w_{i}=\epsilon
_{ijk}\Omega _{j}r_{k}$. Here ${\bf \Omega }$ is the angular velocity and
\begin{eqnarray}
I &=&\frac{4\pi }{r^{3}}\int_{0}^{R}dr^{\prime }r^{\prime \;2}\left(
r^{2}-r^{\prime \;2}\right) \theta \left( r-r^{\prime }\right) \rho \left(
r^{\prime }\right) \,, \\
J &=&\frac{4\pi }{3}\int_{0}^{R}dr^{\prime }\left[ \frac{r^{\prime 4}}{r^{3}}
\theta \left( r-r^{\prime }\right) +r^{\prime \;}\theta \left( r^{\prime
}-r\right) \right] \rho \left( r^{\prime }\right) \,.
\end{eqnarray}

In presence of VEP all the PPN parameters can depend on the flavor numbers.
We assume that deviations from a metric theory are small, so that in a very
good approximation there is a common coordinate frame for all flavors. Since
the parameters are flavor dependent, distinct neutrinos will undergo
different phase shifts when passing through the same sector of space. In the
presence of neutrino mixing phase shift differences become observable as
neutrino oscillations. For simplicity, in what follows we consider two
neutrino flavors, $\nu _{e}$ and $\nu _{\mu }$ or $\nu _{\tau }$. They are
supposed to be linear superpositions of the gravitational eigenstates $\nu
_{1}^{g}$and $\nu _{2}^{g}$, with a mixing angle $\theta _{g}$. Along the
neutrino path flavor evolution is governed by
\begin{equation}
i\frac{d}{dr}\left(
\begin{tabular}{l}
$\nu _{e}$ \\
$\nu _{\mu }$
\end{tabular}
\right) =\frac{\Delta _{0}}{2}\left(
\begin{tabular}{ll}
-cos$2\theta _{g}$ & $\sin 2\theta _{g}$ \\
\ $\sin 2\theta _{g}\,$ & cos$2\theta _{g}$
\end{tabular} \right) \left(
\begin{tabular}{l}
$\nu _{e}$ \\
$\nu _{\mu }$
\end{tabular}
\right) \,,  \label{alfa}
\end{equation}
with $\Delta _{0}=E^{2}-E^{1}$. For a rotating protoneutron star we have
\begin{eqnarray}
\Delta _{0} &=&\left\{ -(\delta \gamma ^{\prime }+\delta \gamma )U-\delta
\Gamma J-\delta \Gamma I({\bf \hat{r}\cdot \hat{p})}^{2}\right.  \nonumber \\
&&+\left[ (\delta \alpha _{2}-\frac{1}{2}\delta \alpha _{1})U-\delta \alpha
_{2}J\right] {\bf v\cdot \hat{p}}-\delta \alpha _{2}I({\bf \hat{r}\cdot v})
{{\bf \hat{r}\cdot \hat{p}}}  \nonumber \\
&&\left. -\frac{1}{2}\left( 7\delta \Delta _{1}+\delta \Delta _{2}\right) J
{\bf \Omega }\times {\bf r\cdot \hat{p}}\right\} E\,,  \label{delta}
\end{eqnarray}
where $E=p$ is the neutrino energy, $\delta \gamma =\gamma ^{2}-\gamma ^{1}$,
and the same for the difference between the other PPN parameters. Here
$\Delta_{0}$ plays the same role as the quantity $(m_{2}^{2}-m_{1}^{2})/2E$
in the mass mechanism for neutrino oscillations. Note that in our case the
potentials depend on $r$ and hence $\Delta _{0}=$ $\Delta _{0}(r)$. Terms
with ${\bf v}$ appear whenever a preferred frame exists. In principle ${\bf
v }$ could also depend on the gravitational flavor, but the observed
position offset for pulsar-supernova remnant pairs\cite{polarization} can be
interpreted as the existence of a translational effect associated to a
preferred direction. For this reason we take ${\bf v}$ as a flavor
independent parameter. Its action is analogous to the one produced by a
magnetic field in the KS mechanism.

As is well known, neutrino oscillations in matter differ from the
oscillations in vacuum. The interaction of neutrinos with the background
modifies their dispersion relations, and under favorable conditions leads to
the MSW phenomena of resonant flavor transformation. If electrons are the
only leptons present in the medium, the term $\frac{G_{F}}{\sqrt{2}}
N_{e}(r)\sigma _{3}$ has to be added to the matrix in Eq.(\ref{alfa}), where
$\sigma _{3}$ is the Pauli matrix and $N_{e}(r)$ denotes the electron number
density. The resulting Hamiltonian can be diagonalized at every point by a
local rotation, with the mixing angle in matter $\theta _{m}(r)$ given by
${\rm \sin }2\theta _{m}(r)=\frac{\Delta _{0}(r)\ }{\Delta (r)}{\rm \sin }
2\theta _{g}$, with
\begin{equation}
\Delta (r)=\sqrt{\left( \Delta _{0}(r)\ {\rm \cos }2\theta _{g}-\sqrt{2}
G_{F}N_{e}(r)\right) ^{2}+\left( \Delta _{0}(r)\ {\rm \sin }2\theta
_{g}\right) ^{2}}\,.
\end{equation}

There is a resonance when the diagonal elements of the Hamiltonian vanish,
i.e. when $\sqrt{2}G_{F}N_{e}(r_{R})=\Delta _{0}(r_{R})\,\cos 2\theta _{g}$.
The efficiency of the flavor transformation depends on the adiabaticity of
the process, which is characterized by the parameter
\begin{equation}
\kappa =\left| {\displaystyle {\frac{1 }{\Delta }}}{\displaystyle {\frac{
d\theta _{m} }{dr}}}\right| _{r=r_{R}}=\left| \Delta _{0}\frac{\sin 2\theta
_{g}\tan 2\theta _{g} }{h_{N_{e}}^{-1}-h_{\Delta _{0}}^{-1}}\right|
_{r=r_{R}} \,.  \label{adiaba}
\end{equation}
where the scale heights are $h_{N_{e}}^{-1}=\frac{d}{dr}\ln N_e$ and
$h_{\Delta _{0}}^{-1}=\frac{d}{dr}\ln \Delta _{0}$. The transition will be
adiabatic whenever $\kappa \gg 1$.

The translational kick comes from the anisotropy in the radial momentum
carried by neutrinos emerging from the resonance surface. The resulting
effect on the motion of the pulsar is obtained integrating over all the
surface. To this integration will only contribute the radial component of
${\bf \hat{p}}$. Therefore, to estimate the translational kick we use a
simplified situation with a purely radial neutrino flux. In this case,
\begin{equation}
\Delta _{0}=\left[ A\left( r\right) +B(r)v\cos \chi \right] E \,,
\end{equation}
where $\chi $ is the angle between ${\bf r}$ and ${\bf v}$. The functions
$A(r)$ and $B(r)$ are given by
\begin{eqnarray}
A &=&-(\delta \gamma ^{\prime }+\delta \gamma )U-\delta \Gamma \left(
I+J\right)\,, \\
B &=&(\delta \alpha _{2}-\frac{1}{2}\delta \alpha _{1})U-\delta \alpha
_{2}\left( I+J\right)\,.
\end{eqnarray}

The radius of a point on the distorted resonance surface can be written as
$r_{R}=r_{o}+\delta $ cos$\chi$ ($\delta \ll r_{o}$). The radius of the
unperturbed resonance sphere $r_{o}$ is determined by
\begin{equation}
A\left( r_{o}\right) =\frac{\sqrt{2}G_{F}}{\cos 2\theta _{g}}\frac{
N_{e}(r_{o})}{E} \,,
\end{equation}
and
\begin{equation}
\delta =\left.\frac{B}{A}\frac{v}{h_{N_{e}}^{-1}-h_{A}^{-1}}
\right|_{r_{o}}\,,
\end{equation}
where we keep only the terms linear in $\delta$, and $h_{A}^{-1}=
{\displaystyle{\frac{d }{dr}}}\ln A(r)$.

At the moment there is no agreement about the details of the
production of a kick by a distorted neutrinosphere. To explore the
possibilities of the VEP mechanism we will now consider this
effect in the context of the main neutrinosphere models proposed.

For a hard neutrinosphere model in thermal equilibrium as considered in Refs.
\cite{kusegre,kusegref,qian}, the momentum asymmetry in the ${\bf v}$
direction is generated by the emission at points with different temperatures
on the resonance surface: $\Delta p/p\approx \frac{2}{9}h_{T}^{-1}\delta $,
where $h_{T}^{-1}=\frac{d}{dr}\ln T $. In the case of a quasi-degenerate gas
of relativistic electrons with a constant chemical potential $\mu _{e}\approx
\left( 3\pi ^{2}N_{e}\right) ^{1/3}$and $\frac{dN_{e}}{dT}=\frac{2}{3}T\mu
_{e}$. Then
\begin{equation}
\frac{\Delta p}{p}\approx Q\frac{Bv}{A}\,,  \label{asym}
\end{equation}
with $Q=\frac{\eta ^{2}\Lambda }{9\pi ^{2}}$, where $\eta =\mu _{e}/T$ is
the degeneracy parameter for the electrons and $\Lambda
=h_{A}/(h_{A}-h_{N_{e}})$. Another possibility is to assume that the
electron fraction $\ Y_{e}$ remains constant and $\rho \sim
T^{3}$\cite{qian}. In this case $h_{N_{e}}\sim h_{T}/3$, and
$Q=\frac{2}{27}\Lambda $.

A different kick model in the literature uses a soft neutrinosphere\cite
{kusegref,raffelt}. In such a case there is an important reduction in the
anisotropy given by the ratio ${\rho _{o}}/{\rho _{c}}$ of the density at
the resonance and the density at the core. The momentum asymmetry can also
be written as in Eq.(\ref{asym}), with $Q=\rho _{o}h_{N_{e}}\Lambda
/18m_{c}$, where $m_{c}= \int_{r_{c}}^{r_{s}}\rho \,dr$ is the integral of
the mass density between the central core and the surface of the star. In
all the cases considered above the adimensional parameter $Q$ depends only
on the specific model and the remaining factors contain the PPN parameters.

For a quantitative estimation of the effects of VEP on the neutrinosphere,
we use the density profile $\rho (r)=\rho _{c}$ for $r<r_{c}$ and $\rho
(r)=\rho _{c}\left( r_{c}/r\right) ^{n}$ for $r>r_{c}$\cite{horvat}. We take
$\rho _{c}=8\times 10^{14}\;g/cm^{3}$, $r_{c}=10\;km$, and $5\leq n\leq 7$
that give a good description of the supernova SN1987A\cite{parametros}. The
resonance surface has to lie below the $\nu _{e}$ neutrinosphere and above
the $\nu _{\mu }$ neutrinosphere. If we take $\rho _{o}\sim 10^{-11}g/cm^{3}$
and $Y_{e}\sim 0.1$, then we obtain $\left( \delta \gamma +\delta \gamma
^{^{\prime }}+0.95\,\delta \Gamma \right) \cos 2\theta _{g}\simeq -6\times
10^{-10}$. For $\delta \Gamma =0$ our result agrees with the one obtained in
Ref.\cite{horvat}. As pointed out in this work the adiabaticity condition is
achieved provided that $\theta _{g}>10^{-4}$, $h_{N_{e}}^{-1}\lesssim
h_{\Delta_{0}}^{-1}$, and hence $\Lambda \simeq 1$ for every value of $n$.
The value of the momentum asymmetry is
\begin{equation}
\frac{\Delta p}{p}\simeq -Q\left( \delta \alpha _{1}-0.1\delta \alpha
_{2}\right) v\cos 2\theta _{g}\times 10^{9}\,.
\end{equation}
For $T=3$MeV, $Q\sim 0.1$ in the hard neutrinosphere models, and $Q\simeq
4\times 10^{-5}$ for a soft neutrinosphere. Taking $\delta \alpha _{1}\sim
\delta \alpha _{2}\sim \delta \alpha $, and requiring $\Delta p/{p}\sim
0.01$, we obtain $v\delta \alpha\sim 10^{-10}$ and $v\delta\alpha \sim
10^{-7}$, respectively.

We now analyze the effect of the non radial component of the neutrino
momentum. When ${\bf \Omega }=0$, at a given point of the resonance surface
the emitted neutrinos have an azimuthal symmetry respect to the position
vector. For non vanishing angular velocity of the protoneutron star, the
last term in Eq. (\ref{delta}) brakes this symmetry and produces an angular
acceleration of the star. To make a perturbative estimation of this effect
we ignore the dependence of $\Delta _{o}$ on $v$ and adopt a very simple
model of a hard resonance surface at $r_{0}+\delta r$. From the resonant
condition we get
\begin{equation}
\delta r=\left. \frac{C}{A}{\displaystyle {\Lambda h_{N_{e}}}}\;{\bf \Omega
\cdot r\times \hat{p}}\right| _{r_{o}}\,,
\end{equation}
where $C(r)=-\frac{1}{2}\left( 7\delta \Delta _{1}+\delta \Delta _{2}\right)
J\left( r\right) $.

Neutrinos emitted in different directions come from regions at different $r$
and therefore at different energies. Hence they have different angular
momenta. If we adopt the Stefan-Boltzmann law for the neutrino flux at the
resonance surface, a neutrino emitted in a direction ${\bf \hat{p }}$ has a
momentum $p=E_{o}(1+4h_{T}^{-1}\delta r)$, where $E_{o}=E(r_{o})$. Therefore
it carries an angular momentum
\begin{equation}
{\bf l=}r_{o}E_{o}({\bf \hat{r}}\times {\bf \hat{p})}\left[
1+4h_{T}^{-1}\delta r\right] \,.
\end{equation}
By integrating at each point of the resonance surface over all directions
and also over all the points, we compute the angular momentum gained by the
star. Because of the symmetry of the system the resulting angular
acceleration points along the rotational axis. The time derivative of the
total angular momentum can be expressed as
\begin{equation}
\dot{L}=\frac{C\Lambda }{3\pi A}\frac{h_{N_{e}}}{h_{T}}{\dot{{\cal E}}}
\Omega r_{o}\frac{\int_{0}^{\pi }d\theta \sin \theta\int_{0}^ {\frac{\pi }{2}
}d\theta^{\prime }\int_{0}^{2\pi }d\varphi ^{\prime } \sin \theta ^{\prime
}r_{o}\left( {\bf \hat{\Omega}}\times {\bf \hat{r}\cdot \hat{p}} \right)
^{2} }{\int_{0}^{\pi }d\theta \sin \theta \int_{0}^{\frac{\pi }{2} }d\theta
^{\prime }\sin \theta ^{\prime }}\,,  \label{L}
\end{equation}
where ${\dot{{\cal E}}}$ is the energy carried by the neutrinos per time
unit, and a factor $\frac{1}{6}$ has been included to take into account that
although six neutrino and antineutrino species are radiated, only one comes
from the distorted neutrinosphere. In the latter expression $\theta $ is the
angle between the radius vector ${\bf \hat{r}}$ and the angular velocity
${\bf \Omega }$, while $\theta ^{\prime }$ and $\varphi ^{\prime }$ are the
spherical coordinates for ${\bf \hat{p}}$ taking ${\bf \hat{r}}$ as the $z$
axis. From Eq. (\ref{L})
\begin{equation}
\Omega (t)=\Omega _{o}\exp \left( \frac{4r_{o}^{2}}{27}\int_{t_{0}}^{t}\frac{
C\Lambda }{AI}\frac{h_{N_{e}}}{h_{T}}{\dot{{\cal E}}}dt\right) \,,
\end{equation}
where $I$ and $\Omega _{o}$ are the momentum of inertia and the initial
angular velocity of the protostar. It should be noted that the rotational
kick does not require a velocity ${\bf v}$ associated to a preferred frame.

As an example, let us consider the density profile introduced above.
Assuming that all the quantities in the integrand except ${\dot{{\cal E}}}$
are constant during the cooling period and taking $\Delta {\cal E}\sim
3\times 10^{53}erg$, the angular velocity after the angular kick is
\begin{equation}
\Omega _{f}\simeq \Omega _{o}\exp \left[ \xi \left( \delta \Delta _{1}+\frac
{1}{7}\delta \Delta _{2}\right) \frac{h_{N_{e}}}{h_{T}}\;\,10^{8}\right] \,,
\end{equation}
where $0.1<\xi <10$ for $5<n<6$. The ratio $\frac{h_{N_{e}}}{h_{T}}$ is a
model dependent quantity of the order of unity. Values usually considered
are $h_{N_{e}}\lesssim h_{T}/3$. The star angular velocity will increase or
decrease depending on the sign of $\delta \Delta _{1}+\frac{1}{7}\delta
\Delta _{2}$. If we accept that typical initial angular velocities of the
protostar are $\Omega _{o}\sim 0.01\Omega _{f}$, the VEP parameters must be
in the range $10^{-6}\;\lesssim \delta \Delta _{1}+\frac{1}{7}\delta \Delta
_{2}\lesssim 10^{-8}$ to reproduce the observed values for $\Omega _{f}$.

To estimate the order of magnitude of the translational and rotational
accelerations, we have assumed the corresponding kicks decoupled one from
the other. In a more realistic situation the rotational motion could produce
an average in the translational kick. This effect depends on the relation
between the characteristic time of reaccommodation of the neutrinosphere and
the period of rotation of the star. The anisotropy axis coincides at every
time with ${\bf v}$ and is not affected by the rotation, but the temperature
of the resonance surface could change. In a soft neutrinosphere the deformed
resonance surface changes the atmosphere opacity over the core and induces a
temperature anisotropy in the core-atmosphere interface, which in turn
affects the neutrino flux. As the star rotates the resonance surface also
rotates with respect to the rest frame of the star inducing a time-changing
opacity over the core region. Therefore we have here to consider the
characteristic thermal response time of the system, which is of the order of
a few hundred miliseconds\cite{raffelt}, in contrast with the pulsar period.
Thus, in this case we can expect an averaging of the translational kick.
This effect tends to cancel the component orthogonal to the rotational axis
and develops a correlation between the translational kick and the axis of
rotation. In the case of a hard neutrinosphere the energy flux depends on
the temperature at the point from which the neutrinos are radiated from the
surface of resonance. The atmosphere here has enough heat capacity to act as
a thermal reservoir with a radius dependent temperature. Therefore, there is
no effective average and there is no correlation between the translational
kick and the rotational axis.

For simplicity, we have assumed that the only mechanism responsible for the
pulsar motion is VEP. If this were the only cause for the translational
velocity, then all pulsar velocities should show a certain correlation
driven by the ${\bf v}$ parameter. This correlation will be more or less
accentuated depending on how hard or soft the neutrinospheres are, and also
could be blurred by the presence of other kick mechanisms besides the one
here considered.

In conclusion, we have shown that resonant VEP neutrino oscillations may be
responsible for both the translational and rotational motion of pulsars.
Since this mechanism works even for massless neutrinos, it does not clash
with cosmological bounds. The strictest boundaries known at present in the
neutrino sector are given by accelerator experiments, mainly from CCFR, which
correspond to the highest tested energies\cite{ccfr}. These experiments are
sensitive to large mixing angles, because they have no access to the MSW
effect. The exclusion region for these experiments extends down to $\sin^2
2\theta>2.10^{-3}$ for $\nu_e\-nu_\mu$ and $\sin^2 2\theta>0.2$ for
$\nu_e\-nu_\tau$, independently of the value of $\Delta_0$. Therefore, the
parameter region relevant for the neutrino resonance in neutron stars,
taking $ 10^{-4}<\theta _{g}<10^{-3}$ for $\nu_e\-nu_\mu$ or $
10^{-4}<\theta _{g}<10^{-1}$ for $\nu_e\-nu_\tau$, is well outside the range
tested by accelerators\cite{mann}. With respect to atmospheric neutrinos,
they are not affected by these small mixing angle oscillations, and the MSW
effect in the solar neutrinos corresponds to a medium of much lesser
density, and thus the involved parameter sector is very
different\cite{nuestro}. In this way the kick pulsar physics gives access to
a new phenomenological sector of VEP effects.

\section{Acknowledgments}

This work was partially supported by CONICET-Argentina, CONACYT- M\'{e}xico,
Universidad Nacional Aut\'{o}noma de M\'{e}xico under grants DGAPA-IN117198
and DGAPA-IN100397, and Centro Latino Americano de F\'{\i}sica. M. B. also
acknowledges support from SRE (M\'{e}xico).

\end{document}